\documentclass[journal]{IEEEtran}
\usepackage{graphicx}
\usepackage{cite}
\usepackage{picinpar}
\usepackage{amsmath}
\usepackage{url}
\usepackage{flushend}
\usepackage[latin1]{inputenc}
\usepackage{colortbl}
\usepackage{soul}
\usepackage{multirow}
\usepackage{pifont}
\usepackage{color}
\usepackage{alltt}
\usepackage[hidelinks]{hyperref}
\usepackage{enumerate}
\usepackage{siunitx}
\usepackage{breakurl}
\usepackage{epstopdf}
\usepackage{pbox}
\usepackage{makecell}
\usepackage{amssymb}
\usepackage{algorithm}
\usepackage{algorithmic}
\ifCLASSINFOpdf
\else
\fi
\begin{document}
\title{Driving Conflict Resolution of Autonomous Vehicles at Unsignalized Intersections: A Differential Game Approach}

\author{Peng Hang,
          Chao Huang, Zhongxu Hu, and Chen Lv
       % <-this % stops a space
\thanks{This work was supported in part by A*STAR National Robotics Programme (No. SERC 1922500046), and in part by SUG-NAP Grant (No. M4082268.050), Nanyang Technological University, Singapore.}% <-this % stops a space
\thanks{P. Hang, C. Huang, Z. Hu, and C. Lv are with the School of Mechanical and Aerospace Engineering, Nanyang Technological University, Singapore 639798. (Emails: \{peng.hang, chao.huang, zhongxu.hu, lyuchen\}@ntu.edu.sg)}% <-this % stops a space

\thanks{Corresponding author: C. Lv}}


\markboth{ }%IEEE Transactions on Intelligent Transportation Systems
{Shell \MakeLowercase{\textit{et al.}}: }



\maketitle

\begin{abstract}
Considering personalized driving preferences, a new decision-making framework is developed using a differential game approach to resolve the driving conflicts of autonomous vehicles (AVs) at unsignalized intersections. To realize human-like driving and personalized decision-making, driving aggressiveness is first defined for AVs. To improve driving safety, a Gaussian potential field model is built for collision risk assessment. Besides, in the proposed decision making framework, the collision risk assessment model is further used to reduce the computational complexity based on an event-triggered mechanism. In the construction of payoff function, both driving safety and passing efficiency are comprehensively considered, and the driving aggressiveness is also reflected. Two kinds of equilibrium solution to the differential game, i.e., the Nash equilibrium and Stackelberg equilibrium, are discussed and solved. Finally, the proposed decision making algorithm is tested through a hardware-in-the-loop testing platform, and its feasibility, effectiveness and real-time implementation performance are validated.

\end{abstract}


\IEEEpeerreviewmaketitle

\section{Introduction}
\IEEEPARstart{W}{ith} the effective control of traffic signal lights, the driving conflicts of vehicles at intersections can be resolved [1], and thus the driving safety of vehicles is guaranteed. However, the traffic efficiency is limited due to the control cycle of traffic signals [2]. In the context of intelligent transportation system and vehicle to everything (V2X), the traditional traffic light control becomes less important. Autonomous vehicles (AVs) can make intelligent decisions according to the information of surrounding objects and infrastructures [3,4], and both driving safety and passing efficiency can be further improved [5]. Therefore, the decision making of AVs at unsignalized intersections becomes a research hotpot in recent years.

In most of exisiting studies, the driving conflict resolution of AVs at unsignalized intersections is converted into an optimization problem, which addresses the passing order scheduling and velocity planning of AVs [6]. In [7], the passing order of AVs is optimized with the Monte Carlo tree search (MCTS), which can realize a good balance between performance and computation flexibility. In [8], a three-level distributed control algorithm is proposed to solve the driving conflict at unsignalized intersections. The upper level is used for state observation of AVs, the middle level is designed for scheduling conflict-free arriving time, and the lower level is used for addressing parameter mismatch and acceleration saturation. In [9], a centralized multi-vehicle coordination scheme is proposed, which uses Mixed Integer Linear Programming (MILP) to obtain the passing sequence of AVs. With the centralized traffic management approach, AVs must give up their self control rights and follow the scheduling proposed by the centralized controller [10]. Different from the centralized management concept, a communication-enabled distributed conflict resolution approach is designed, which provides an efficient parallel mechanism to obtain local optimal solutions [11]. The algorithm performance is limited by the amount of vehicles. With the increase of AVs, the solutions may be unacceptable. In [12], a receding horizon model predictive control (MPC) method is utilized to simultaneously optimize the passing sequence of AVs and improve the fuel economy. With the application of MPC, it brings a challenge on the computing efficiency of the algorithm. In [13], the swarm intelligence approach is adopted in  the intelligent decision making framework of AVs at unsignalized intersections. In this framework, each AV is able to do self-calculation and make adaptive decisions following the traffic dynamics. In [14], a virtual platoon is designed to solve the driving conflicts of AVs at unsignalized intersections by constructing the conflict-free geometry topology. Additionally, the comparative study between the ad hoc negotiation-based strategy and the planning-based strategy is conducted in [15]. The results indicate that there is no significant difference on performance in the low traffic flow scenario, but the planning-based strategy shows a better performance in the high flow-rate traffic scenario.

The above studies mainly address the driving conflicts of AVs in the single-lane cross intersection scenario, however, more complex intersection scenarios can hardly be solved. To address driving conflicts at multi-lane intersections, a multi-objective optimal control model is proposed by jointly considering vehicle safety, energy consumption, and ride comfort [16]. In [17], a behavioural decision-making model is constructed for AVs using driving risk assessment. It can realize collision avoidance on the premise of ensuring a certain efficiency at multi-lane intersections. In [18], a scheduling algorithm is designed for AVs at unsignalized intersections with an alternately iterative descent method. In [19], a traffic conflict model is proposed to evaluate the driving safety of AVs at intersections.
To improve the computing efficiency of the decision-making algorithm, especially in the high traffic flow rate scenario, a novel parallel computation framework is proposed with the alternating direction method of multipliers (ADMM). Moreover, it can also address the nonlinear and nonconvex optimization problem [20]. Besides, learning-based approaches have been widely used for autonomous decision making. In [21], an end-to-end decision-making framework is designed by using a convolutional neural network to map the relationship between traffic images and vehicle operations. In [22], a safe reinforcement learning (RL) approach is proposed to address the driving conflicts at autonomously navigate intersections, which shows good robustness against perception errors and occlusions. To deal with the left-turn decision-making issue of AVs, a deep reinforcement learning (DRL) decision-making framework is designed, which can efficaciously reduce the collision rate and improve transport efficiency [23]. In [24], a nondominated sorting genetic algorithm (NSGA-II) and deep deterministic policy gradient (DDPG) are combined to deal with the driving conflicts of AVs at complex and dynamic urban intersections.
In addition, the game-theoretic approach shows superiority and effectiveness on the modelling of AVs' interaction and decision-making [25]. Focusing on unprotected left turn maneuvers, a decision-making framework is built for AVs with game theory in [26], but only two players are discussed in the framework. In [27], multi-player general-sum differential game is used in the decision-making algorithm for AVs, and it realizes the interaction and conflict resolution of multi-vehicle at unsignalized intersections. In [28], the differential game approach is used to realize collision avoidance for multi-agent. In [29], to improve the decision-making performances of AVs, the noncooperative game theory is combined with Q-learning to solve the driving conflicts of AVs.
In general, learning-based approaches are primarily driven by data. To this end, their performances are limited by the quality of the dataset. Besides, the learning-based approach has poor interpretability. Once there exits an error, it is difficult to find the cause and propose a solution. However, game theoretic approaches show advantages in the interaction modelling and the decision-making behaviour analysis. Namely, human-like decision making and driving are easy to achieve.

The above literature review shows that different kinds of approaches have been proposed to address the driving conflicts of AVs at unsignalized intersections. However, the personalized driving demands are seldom discussed in the existing studies. On one hand, AVs will be deployed with different purposes in the future, e.g., ambulance, police car, fire truck, school bus, etc. And different types of vehicles have various demands on safety, performance, fuel economy, comfort, etc [30].
On the other hand, different passengers have various individual travel demands. For instance, pregnant women and elderly people would prefer a higher level on driving safety and riding comfort, but daily commuters hope to shorten their travel time during peak hours [31]. Therefore, personalized demands and features are of great importance for future AV design. Currently, most of the studies consider AVs with a same normal driving style under the unsignalized intersection scenario, and the effects of AVs' driving characteristics on the decision-making behaviors have rarely been reported.

To further advance the technology and solve the driving conflicts of AVs at unsignalized intersections, in this paper, a novel decision-making framework is proposed. The contributions are summarized as follows: (1) The driving aggressiveness, which can help AVs realize human-like driving and personalized decision making at unsignalized intersections, is defined for AVs; (2) A collision risk assessment algorithm is proposed with the a Gaussian potential field approach and an event-triggered mechanism, which can not only guarantee the driving safety of AVs, but also effectively improve computation efficiency of the decision-making algorithm; (3) The differential game approach is used to model the human-like interaction and decision making of AVs, addressing the driving conflicts of AVs at the multi-lane unsignalized intersection.

The organization of the remaining content is described as follows. The problem formulation of the driving conflict resolution for AVs at unsignalized intersections is presented in Section II. In Section III, the vehicle model and collision risk assessment model are developed. Then, the differential game approach is applied to solve the decision-making of AVs at the unsignalized intersection in Section IV. In Section V, the algorithm is validated on a hardware-in-the-loop testing platform. Finally, conclusions are presented in Section VI.

\section{Problem Formulation}

\begin{figure}[t]\centering
	\includegraphics[width=8.5cm]{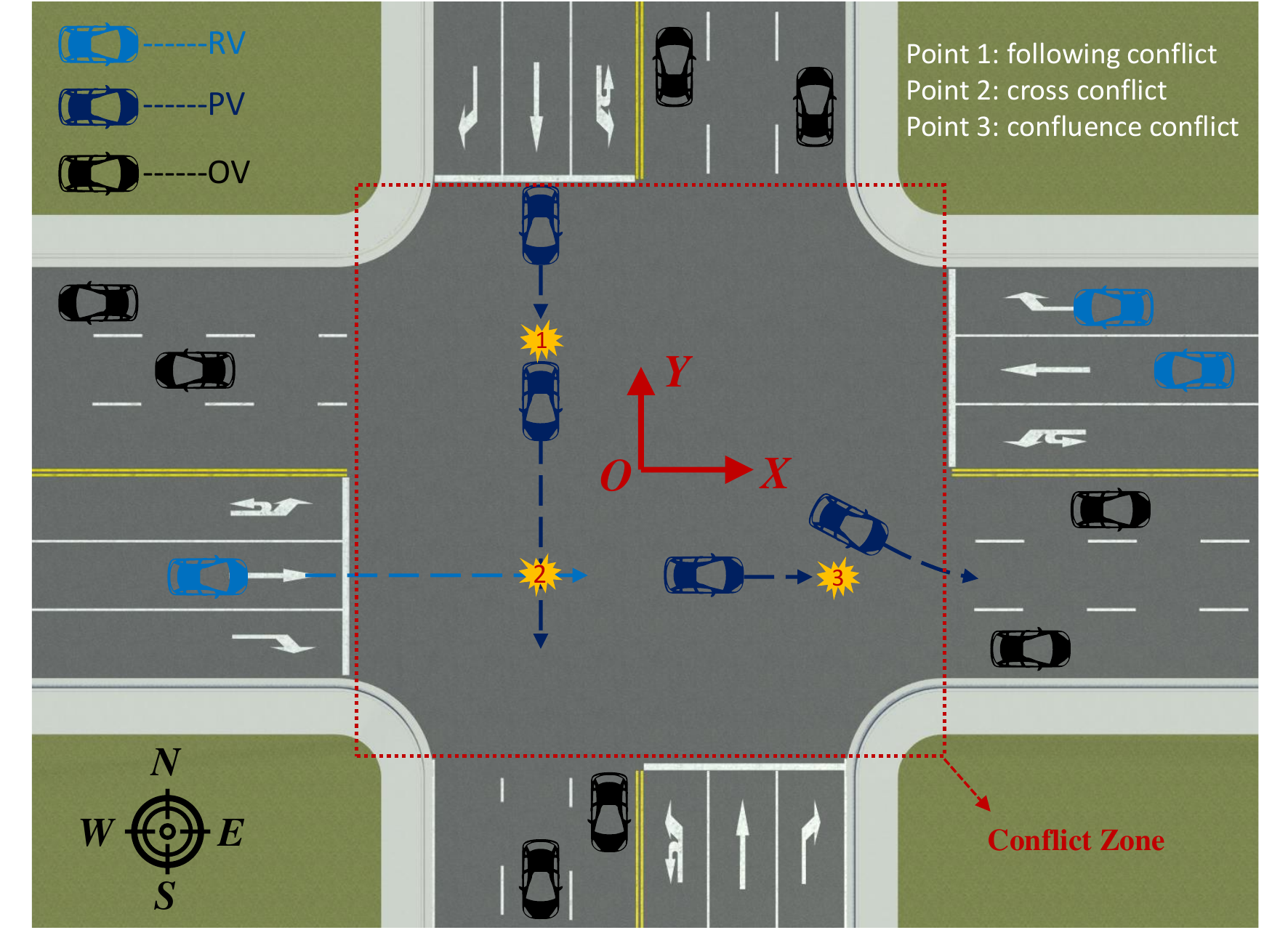}
	\caption{Unsignalized intersection scenario.}\label{FIG_1}
\end{figure}

Fig. 1 shows the unsignalized intersection scenario, in which all vehicles are assumed to be AVs. The unsignalized intersection, i.e., the conflict zone (CZ), is linked with eight three-lane roads. According to the positions of AVs in the CZ, all vehicles are divided into three types: ready vehicle (RV), which is ready to enter the intersection; passing vehicle (PV), which is passing the intersection; outside vehicle (OV), which has moved out of the intersection. In Fig. 1, RV, PV and OV are marked as blue vehicle, purple vehicle and black vehicle, respectively. Only RV and PV are discussed in the driving conflict issue of AVs at unsignalized intersections. Additionally, from the perspective of individual decision making, all vehicles can be divided into four kinds according to the relative relationship and driving conflict, i.e., host vehicle (HV), leading vehicle (LV), neighbour-conflict vehicle (NV), and irrelevant vehicle (IV). Each vehicle can be regarded as HV. The vehicle that moves in front of HV is called LV. If the vehicle's motion direction is different from HV and has a conflict risk with HV, it is called a NV. Except LV and NV, other vehicles are regarded as IV for HV. The names of NV, LV and IV are relative and updated according to the driving situation in real time.

In the unsignalized intersection scenario, according to the driving directions and moving trajectories of two competitor vehicles, three kinds of driving conflict are defined, i.e., the following conflict, cross conflict and confluence conflict.

1. Following conflict: If two vehicles have the same driving direction and moving trajectory, and the velocity of HV is larger than that of LV, there will be a following conflict occurring between HV and LV.

2. Cross conflict: If two vehicles have different driving directions and moving trajectories, and there exists at least one intersection of the two trajectories, then it is called a cross conflict.

3. Confluence conflict: If vehicles want to merge into the same lane from different lanes, there will be a confluence conflict.

The conflict points (CPs) shown in Fig. 1 illustrate the three types of driving conflict for AVs at unsignalized intersections.
As mentioned in Introduction, the driving aggressiveness has significant effect on the decision-making behavior. Thus it is also considered in the decision-making problem of AVs.

To address the driving conflicts of AVs at unsignalized intersections, a novel decision-making framework is designed and presented in Fig. 2. It mainly consists of two modules, i.e., modelling, and decision-making algorithm. In the modelling module, a vehicle kinematic model is used for collision prediction and driving risk assessment. The collision risks of AVs are assessed using the potential field approach. The driving aggressiveness and the results of collision risk assessment are outputted to the decision-making module. The payoff function is constructed considering driving safety and passing efficiency of AVs. On the basis of the payoff function and constraints, the differential game approach is used to solve the decision-making issue. Further, two equilibrium conditions of the differential game, i.e., Nash equilibrium and Stackelberg equilibrium, are discussed. Finally, the decision-making results are outputted to the motion planning and control modules. The related studies of collision prediction and aggressiveness estimation have been reported in previous works [32,33], and they will not be introduced in detail in this paper.

\begin{figure}[t]\centering
	\includegraphics[width=8.5cm]{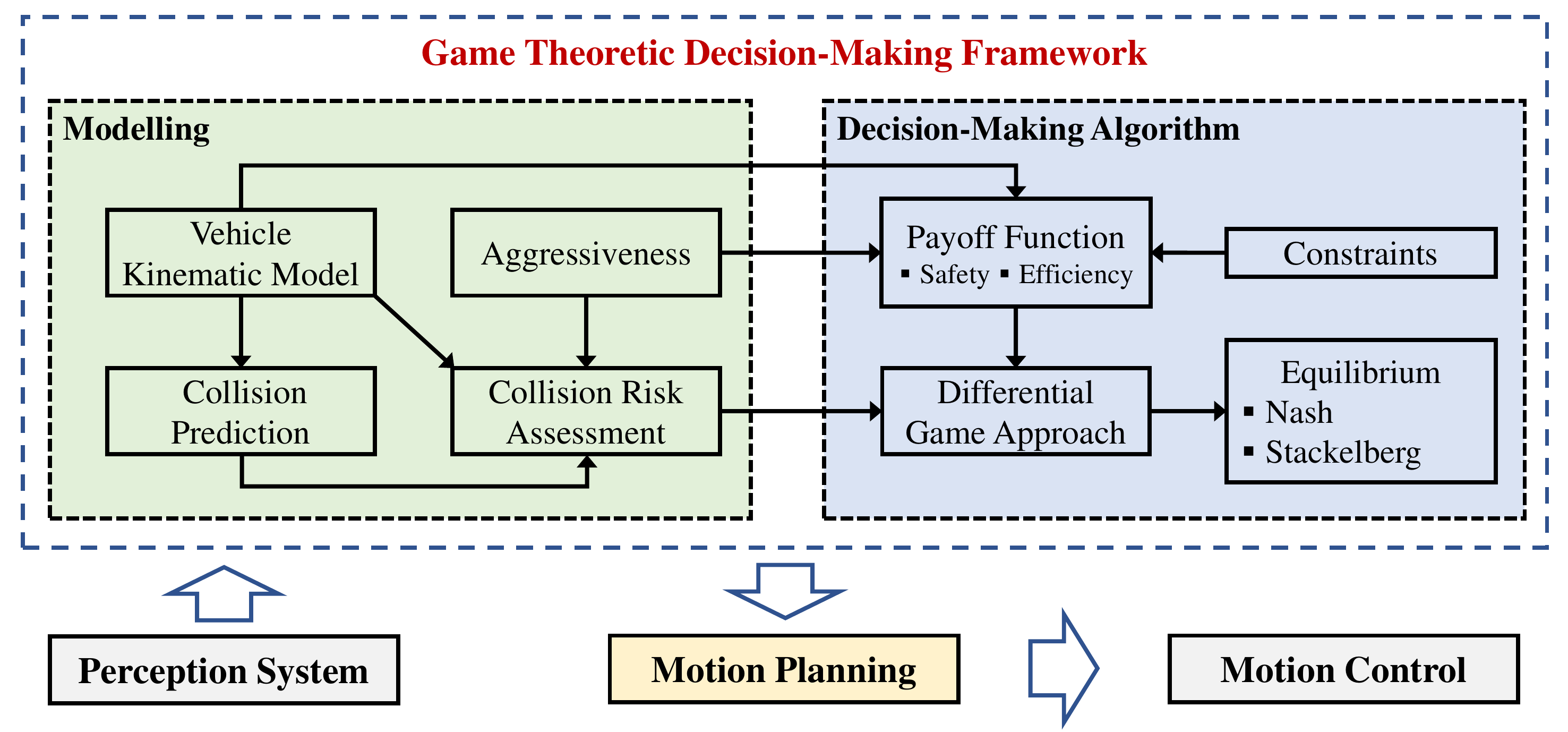}
	\caption{Game theoretic decision-making framework.}\label{FIG_2}
\end{figure}

\section{System Modelling}

\subsection{The Vehicle Model}
To reduce the computational burden, the vehicle kinematic model has been widely applied to the algorithm design of decision making  [34]. In this paper, the simplified bicycle model is utilized. For AV\emph{i}, its kinematic model for decision making is derived.
% Eq.
\begin{align}
\dot{x}^i(t)=f^i(x^i(t),u^i(t))
\end{align}
% Eq.
\begin{align}
f^i(x^i(t),u^i(t))=
&
\left[
\begin{array}{ccc}
a_x^i\\
v_x^i\tan\beta^i/l_r^i\\
v_x^i\cos\phi^i/\cos\beta^i\\
v_x^i\sin\phi^i/\cos\beta^i\\
\end{array}
\right]
\end{align}
% Eq.
\begin{align}
\beta^i=\arctan(l_r^i/(l_f^i+l_r^i)\tan\delta_f^i)
\end{align}
% Eq.
\begin{align}
\Phi^i=\varphi^i+\beta^i
\end{align}
where the state vector and control vector of AV\emph{i} are denoted by $x^i=[v_x^i, \varphi^i, X_g^i, Y_g^i]^{T}$ and $u^i=[a_x^i, \delta_f^i]^{T}$, respectively. $v_x^i$, $\varphi^i$ and $\Phi^i$ are the longitudinal velocity, yaw angle and heading angle of AV\emph{i}, respectively. $(X_g^i, Y_g^i)$  is the coordinate position of the center of gravity. $a_x^i$ and $\delta_f^i$ denote the longitudinal acceleration and the front-wheel steering angle of AV\emph{i}. $\beta^i$ denotes the sideslip angle of AV\emph{i}. $l_f^i$ and $l_r^i$ are the front and rear wheel bases of AV\emph{i}.

\subsection{The Aggressiveness Level}
The aggressiveness level of human drivers can be reflected by some certain driving behaviors, for example, acceleration or deceleration, and steering operations. Aggressive drivers usually prefer a higher passing efficiency, so sudden accelerations and frequent lane change behaviors would occur [35]. On the contrary, conservative drivers often care more about their driving safety [36]. Therefore, a lower driving speed and a large safe gap are usually their common choices, and the passing efficiency will be sacrificed. But in general, driving safety and passing efficiency should always be comprehensively considered, while different drivers may have different preferences and operations to do trade-off.

To realize human-like driving and personalized decision making for AVs, the aggressiveness level, which is defined by $\kappa$, $\kappa\in [0,1]$, is considered  in this work. A larger value of $\kappa$ means a higher passing efficiency and a relatively worse driving safety. On the contrary, a smaller $\kappa$ indicates a lower passing efficiency and a better driving safety.

\subsection{The Collision Risk Assessment Model}
The potential field approach has been widely used in motion planning and driving risk assessment for AVs. In this subsection, a Gaussian potential field model is established for collision risk assessment. The potential field model for AV\emph{i} is described as follows [37].
% Eq.
\begin{align}
\Gamma^i=\Lambda^i\exp (\frac{-(\sqrt{(X-X_c^i)^2+(Y-Y_c^i)^2}-l^i/\tan \delta_f^i)^2}{2(\sigma^i)^2})
\end{align}
% Eq.
\begin{align}
\begin{array}{ccc}
X_c^i=X_g^i-l_r^i\cos\varphi^i+l^i\sin\varphi^i/\tan \delta_f^i\\
Y_c^i=Y_g^i-l_r^i\sin\varphi^i-l^i\cos\varphi^i/\tan \delta_f^i\\
\end{array}
\end{align}
% Eq
\begin{align}
\Lambda^i=a^i(s^i-v_x^i\cdot t_p^i)
\end{align}
% Eq.
\begin{align}
\sigma^i=(b^i+c^i|\delta_f^i|)s^i+d^i
\end{align}
where $(X_c^i, Y_c^i)$ is the position coordinate of turning center. $l^i=l_f^i+l_r^i$, $a^i=a_0^i e^{\kappa^i}$, $a_0^i$ is a fixed value, $\kappa^i$ denotes the aggressiveness of AV\emph{i}, $s^i$ denotes the length of the predicted path, and $t_p^i$ denotes the predicted time. Additionally, $b^i$ defines the slope of widening of the potential field when $\delta_f^i=0$, $c^i$ is the gain coefficient of $|\delta_f^i|$, $d^i=W^i/4$, $W^i$ is the width of AV\emph{i}.

To assess the collision risk of AVs, the collision position should be predicted in advance. To predict the collision position, the motion trajectories of AVs must be predicted at first. In this paper, the trajectory prediction is based on the model-based prediction approach using the kinematic model and polynomial fitting. The detailed algorithm is presented in the previous work [29]. With the predicted trajectories, the collision position $(X_{cp}, Y_{cp})$ can be worked out. Furthermore, the potential field value $\Gamma_{cp}$ at the predicted collision point can be calculated according to the potential field model (5).
% Eq.
\begin{align}
\Gamma_{cp}=\sum\limits_{i=1}^{\varsigma} \Gamma^i|_{(X, Y)=(X_{cp}, Y_{cp})}
\end{align}
where $\varsigma$ denotes the number of AVs that have conflict at the collision position $(X_{cp}, Y_{cp})$. If $\Gamma_{cp}<\Gamma_{sf}$, where $\Gamma_{sf}$ is the defined safe value, it can be considered that there is no risk of collision at this time step.

\section{Decision-Making Algorithm with Differential Game}

\subsection{Description of the Differential Game}
Differential games are a group of problems related to the modelling and analysis of conflict in the context of a dynamical system. More specifically, decision-making variables evolve over time according to the differential equation [38]. At unsignalized intersections, the decision-making behavior of AV\emph{i} is affected by the driving behaviors and motion states of surrounding vehicles. The motion states of vehicles are derived by the kinematic model, i.e., differential equations (1)-(3). The differential equation of vehicle motion is the basis for the application of differential game approaches.

Considering the differential game in continuous time, $t\in[0, T]$, the state equation of the \emph{N-player} decision-making system can be expressed as
% Eq.
\begin{align}
\dot{x}(t)=f(t, x(t), u^1(t), \cdots, u^N(t)); \quad x(0)=x_0
\end{align}
where $x(t)\in \textbf{R}^N$, which includes the states of N players (AVs). The state function $f$ is a combination of $f^1,\cdots, f^N$. $u^i(t)$ is the control vector of AV\emph{i}, which satisfies the following constraint.
% Eq.
\begin{align}
u^i(t)\in U^i; \quad \forall t\in[0, T], \quad i\in \textbf{N}
\end{align}

In the differential game, each player aims to maximum its own payoff. For the \emph{ith} player, i.e., AV\emph{i}, its payoff function is defined by
% Eq.
\begin{align}
P^i\triangleq \phi^i(x(T))-\int_{0}^{T}J^i(t, x(t), u^1(t), \cdots, u^N(t))dt
\end{align}
where $\phi^i$ and $J^i$ denote the terminal payoff and running cost of AV\emph{i}, respectively. To simplify the payoff function, the terminal payoff is defined as a constant value.

\subsection{The Running Cost for Decision Making}
In the running cost function for decision making of AVs, two key driving performance indexes, i.e., driving safety and passing efficiency, are considered. If AV\emph{i} is regarded as HV, the running cost function for AV\emph{i} can be written as
% Eq.
\begin{align}
J^{i}=(1-\kappa^{i})J_s^{i}+\kappa^{i}J_e^{i}
\end{align}
where $J_s^{i}$ and $J_e^{i}$ are the costs of driving safety and passing efficiency. The weights of the two driving performance indexes are expressed with the aggressiveness $\kappa^{i}$. There exists a negative correlation between $\kappa^{i}$ and $J_s^{i}$, and the positive correlation is for $\kappa^{i}$ and $J_e^{i}$.

The cost function of driving safety  $J_s^{i}$ is divided into three parts including longitudinal and lateral safety, and lane-keeping safety.
% Eq.
\begin{align}
J_s^{i}=k_{s-log}^{i} J_{s-log}^{i}+k_{s-lat}^{i} J_{s-lat}^{i}+k_{s-lk}^{i} J_{s-lk}^{i}
\end{align}
where $J_{s-log}^{i}$, $J_{s-lat}^{i}$ and $J_{s-lk}^{i}$ denote the costs of the longitudinal, lateral and lane keeping safety, respectively. $k_{s-log}^{i}$, $k_{s-lat}^{i}$ and $k_{s-lk}^{i}$ denote the weights of the three safety indexes.

$J_{s-log}^{i}$ reflects the longitudinal safety performance of AV\emph{i} in the decision making process, which is defined by the time to collision (TTC) between AV\emph{i} and its LV.
% Eq.
\begin{align}
J_{s-log}^{i}=\gamma_{log}^{i}/ ((TTC_{log}^{i})^2+\varepsilon)
\end{align}
% Eq.
\begin{align}
TTC_{log}^{i}=\Delta s^{i-LV}/\Delta v_{x}^{i-LV}
\end{align}
% Eq.
\begin{align}
\Delta v_{x}^{i-LV}=v_{x}^{i}-v_{x}^{LV}
\end{align}
% Eq.
\begin{align}
\gamma_{log}^{i}=
&
\left\{
\begin{array}{lr}
1,\quad \Delta v_{x}^{i-LV}>0\\
0,\quad \Delta v_{x}^{i-LV}\leq0\\
\end{array}
\right.
\end{align}
where $TTC_{log}^{i}$ denotes the TTC of AV\emph{i} on longitudinal safety. $v_{x}^{LV}$ is the longitudinal velocity of LV. $\Delta v_{x}^{i-LV}$ and $\Delta s_{x}^{i-LV}$ denote the relative velocity and gap between AV\emph{i} and LV. $\varepsilon$ is a constant value, $\varepsilon>0$.

The cost of lateral safety $J_{s-lat}^{i}$ is designed to address the cross and confluence conflicts for AVs.
If NV\emph{j} is the $jth$ NV of AV\emph{i} at the $jth$ CP, the lateral safety cost $J_{s-lat}^{i}$ can be expressed as
% Eq.
\begin{align}
J_{s-lat}^{i}=1/((TTC_{lat}^{i})^2+\varepsilon)
\end{align}
% Eq.
\begin{align}
TTC_{lat}^{i}=TCP^{i}-TCP^{NVj}
\end{align}
% Eq.
\begin{align}
TCP^{i}=\Delta s^{i-CPj}/v_{x}^{i}
\end{align}
% Eq.
\begin{align}
TCP^{NVj}=\Delta s^{NVj-CPj}/v_{x}^{NVj}
\end{align}
where $TTC_{lat}^{i}$ denotes the TTC of AV\emph{i} on lateral safety. $TCP^{i}$ and $TCP^{NVj}$ denote the time to the $jth$ CP of AV\emph{i} and NV\emph{j}.
$\Delta s^{i-CPj}$ and $\Delta s^{NVj-CPj}$ denote the distances from AV\emph{i} to the $jth$ CP, and from NV\emph{j} to the $jth$ CP. $v_{x}^{NVj}$ is the longitudinal velocity of NV\emph{j}.

Additionally, $J_{s-lk}^{i}$ is the cost of the lane-keeping safety for AV\emph{i}, defined with the lateral offset and the error of heading angle.
% Eq.
\begin{align}
J_{s-lk}^{i}=k_{y-lk}^{i}(\Delta y^{i})^2+k_{\varphi-lk}^{i}(\Delta \phi^{i})^2
\end{align}
where $\Delta y^{i}$ and $\Delta \phi^{i}$  are the lateral offset and the error of heading angle. Besides, $k_{y-lk}^{i}$  and $k_{\phi-lk}^{i}$  are the weighting coefficients.

Finally, the cost function of passing efficiency $J_{e}^{i}$ for AV\emph{i} is defined with velocity error between AV\emph{i} and the maximum velocity.
% Eq.
\begin{align}
J_{e}^{i}=k_e^i(v_x^{i}-v_x^{max})^2
\end{align}
where $k_e^i$ is the weighting coefficient, and $v_x^{max}$ denotes the maximum velocity.

\subsection{The Nash Equilibrium of the Differential Game}
In this subsection, we mainly discuss the Nash equilibrium solution to the differential game described in Eqs. 10-12.
The set of strategies $\{u^{1\ast}, \cdots,u^{N\ast}\}$ is defined as a Nash equilibrium of the differential game if the following condition holds.
% Eq.
\begin{align}
\begin{array}{lr}
u^{1\ast}=\arg\underset{u^1}{\max} P^1(u^{1}, u^{2\ast}, \cdots, u^{N\ast})\\
\quad\quad\quad\quad\vdots\\
u^{i\ast}=\arg\underset{u^i}{\max} P^i(u^{1\ast}, \cdots,u^{i-1\ast}, u^{i}, u^{i+1\ast}, \cdots, u^{N\ast})\\
\quad\quad\quad\quad\vdots\\
u^{N\ast}=\arg\underset{u^N}{\max} P^N(u^{1\ast}, \cdots, u^{i-1\ast}, u^{N})\\
\end{array}
\end{align}
subject to % Eq.
\begin{align}
\begin{array}{lr}
\dot{x}(t)=f(t, x, u^{1\ast}, \cdots,u^{i-1\ast}, u^{i}, u^{i+1\ast}, \cdots, u^{N\ast})\\
x(0)=x_0,\quad u^i\in U^i, \quad i\in \textbf{N}
\end{array}
\end{align}

It can be found that, for \emph{N-player} differential game, we need to simultaneously solve \emph{N} optimization problems. If the functions $f$, $\phi^i$ and $J^i$ ($\forall i\in \textbf{N}$) are continuous differentiable, necessary conditions for optimality can be obtained based on the Pontryagin Maximum Principle (PMP) [39].

\textbf{Lemma 1.} For an \emph{N-player} differential game of prescribed fixed duration $[0,T]$, let\\
(a) $\forall t\in [0,T]$, $f(t, x, u^{1}, \cdots, u^{N})$ be continuously differentiable on $\textbf{R}^N$.\\
(b) $\forall t\in [0,T]$, $\forall i\in \textbf{N}$, $J^i(t, x, u^{1}, \cdots, u^{N})$ and $\phi^i$ be continuously differentiable on $\textbf{R}^N$.\\
Then, if $\{u^{1\ast}, \cdots,u^{N\ast}\} \in U^1\times \cdots \times U^N$ is an open-loop Nash equilibrium solution, there exist $N$ costate functions $\xi^i$, $i\in \textbf{N}$, such that
% Eq.
\begin{align}
\left\{
\begin{array}{lr}
\dot{x^{\ast}}=f(t, x^{\ast}, u^{1\ast}, \cdots, u^{N\ast})\\
\dot{\xi^i}=-\xi^i \frac{\mathrm{\partial}f}{\mathrm{\partial}x} (t, x^{\ast}, u^{1\ast}, \cdots, u^{N\ast})+\\   \quad\quad\quad \quad \quad\quad\quad\quad \frac{\mathrm{\partial}J^i}{\mathrm{\partial}x} (t, x^{\ast}, u^{1\ast}, \cdots, u^{N\ast})\\
x^{\ast}(0)=x_0, \quad \xi^i(T)=\nabla\phi^i(x^{\ast}(T)), \quad i\in \textbf{N}
\end{array}
\right.
\end{align}
% Eq.
\begin{align}
\begin{array}{lr}
u^{i\ast}=\arg\underset{u^{i}\in U^i}{\max} \{ \xi^i f(t,x^{\ast},u^{1\ast},\cdots,u^{i-1\ast},u^{i}, u^{i+1\ast},\cdots,u^{N\ast})\\
 \quad\quad\quad \quad -J^i(t,x^{\ast},u^{1\ast},\cdots,u^{i-1\ast},u^{i}, u^{i+1\ast},\cdots,u^{N\ast})\}\\
\end{array}
\end{align}

\emph{Proof}. Consider the $ith$ equation of (25), which says that $u^{i\ast}$ maximizes $P^i(u^{1\ast}, \cdots,u^{i-1\ast}, u^{i}, u^{i+1\ast}, \cdots, u^{N\ast})$ over $U^i$ subject to the state equation (26). This is a standard optimal problem for the $ith$ player since $u^{j\ast}$ $(j\neq i)$ are open-loop controls and hence do not depend on $u^i$. Therefore, the result follows directly from PMP for the control system.

Let $u^{i\ast}(t)$ be an optimal control and $x^{\ast}(t)$ be the corresponding optimal trajectory of the maximization problem (10)-(12), (25). According to the definition of PMP in [40], the vector $\xi^i$ is defined as the solution to the linear adjoint system
% Eq.
\begin{align}
\begin{array}{lr}
\dot{\xi^i}=-\xi^i \frac{\mathrm{\partial}f}{\mathrm{\partial}x} (t, x^{\ast}, u^{1\ast}, \cdots, u^{N\ast})+\\   \quad\quad\quad \quad \quad\quad\quad\quad \frac{\mathrm{\partial}J^i}{\mathrm{\partial}x} (t, x^{\ast}, u^{1\ast}, \cdots, u^{N\ast})\\
\end{array}
\end{align}
with the terminal condition $\xi^i(T)=\nabla\phi^i(x^{\ast}(T))$.

Then, $\forall t\in [0,T]$, the following maximality condition holds,
% Eq.
\begin{align}
\begin{array}{lr}
\xi^i f(t,x^{\ast},u^{1\ast},\cdots,u^{i-1\ast},u^{i\ast}, u^{i+1\ast},\cdots,u^{N\ast})\\
 \quad\quad\quad \quad -J^i(t,x^{\ast},u^{1\ast},\cdots,u^{i-1\ast},u^{i\ast}, u^{i+1\ast},\cdots,u^{N\ast})\\
 =\underset{u^{i}\in U^i}{\max} \{ \xi^i f(t,x^{\ast},u^{1\ast},\cdots,u^{i-1\ast},u^{i}, u^{i+1\ast},\cdots,u^{N\ast})\\
 \quad\quad\quad \quad -J^i(t,x^{\ast},u^{1\ast},\cdots,u^{i-1\ast},u^{i}, u^{i+1\ast},\cdots,u^{N\ast})\}\\
\end{array}
\end{align}

Based on above equation, (28) is derived, i.e., Lemma 1 is proven.

One class of differential games, for which the necessity condition of Lemma 1 is satisfied, is that with those weakly coupled players. And it can be described using the following state equations and payoff functions (taking two players as an instance without any loss of generality):
% Eq.
\begin{align}
\begin{array}{lr}
\dot{x}^i(t)=f^i(t, x^i(t), u^i(t))+\epsilon f^{ij}(t, x^j(t)); \quad x^i(0)=x_0^i\\
\dot{x}^j(t)=f^j(t, x^j(t), u^j(t))+\epsilon f^{ji}(t, x^i(t)); \quad x^j(0)=x_0^j\\
\end{array}
\end{align}
and payoff functions:
% Eq.
\begin{align}
\begin{array}{lr}
P^i=-\int_0^T{[J^{ii}(t, x^i(t), u^i(t))+\epsilon J^{ij}(t, x^j(t),u^j(t))]dt}\\
\quad\quad\quad \quad+\phi^{ii}(x^i(T))+\epsilon \phi^{ij}(x^j(T))\\
P^j=-\int_0^T{[J^{jj}(t, x^j(t), u^j(t))+\epsilon J^{ji}(t, x^i(t),u^i(t))]dt}\\
\quad\quad\quad \quad+\phi^{jj}(x^j(T))+\epsilon \phi^{ji}(x^i(T))\\
\end{array}
\end{align}
where $\epsilon$ denotes a sufficiently small scalar. Under some appropriate convexity and differentiability conditions, it shows that there exists an $\epsilon_0>0$ such that for all $\epsilon\in (-\epsilon_0,\epsilon_0)$, the differential game admits an unique open-loop Nash equilibrium solution that is stable with respect to Gauss-Seidel or Jacobi interactions [41]. Taking the player $i$ as an example, the solution can be obtained by expanding the state and control corresponding to the open-loop Nash equilibrium solution in power series in terms of $\epsilon$,
% Eq.
\begin{align}
\begin{array}{lr}
x^{i\ast}(t;\epsilon)=\sum_{k=0}^\infty x_k^i(t)\epsilon_k, \quad u^{i\ast}(t;\epsilon)=\sum_{k=0}^\infty u_k^i(t)\epsilon_k
\end{array}
\end{align}

Substituting these into equation (28), along with a similar expansion for $\xi^i$, the different terms $x_k^i$ and $u_k^i$,$(k=0,1,\cdots)$ can be solved iteratively. The solving process is similar for player $j$. It turns out that $u_0^i$ and $u_0^j$ are the open-loop optimal controls associated with the decoupled optimal control problem:
% Eq.
\begin{align}
\begin{array}{lr}
\dot{x}^i(t)=f^i(t, x^i(t), u^i(t)), \quad x^i(0)=x_0^i\\
P^i=-\int_0^T{J^{ii}(t, x^i(t), u^i(t))dt}+\phi^{ii}(x^i(T))\\
\end{array}
\end{align}

Therefore, this approach decomposes the original two-player differential game into two nonlinear optimal control problems (the zeroth-order problems) and a sequence of iteratively constructed linear-quadratic control problem. Halting this iteration at the $k$th step yields an $\epsilon^k$-approximate open-loop Nash equilibrium solution.

\begin{figure}[h]\centering
	\includegraphics[width=8cm]{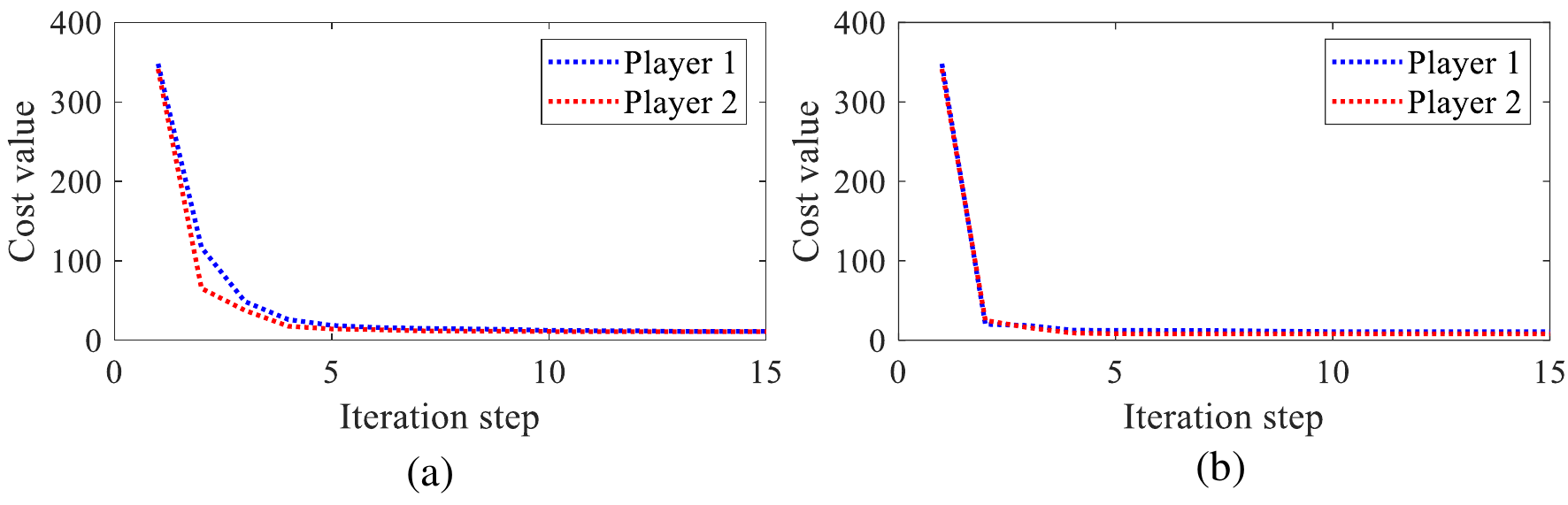}
	\caption{Convergence analysis: (a) Nash equilibrium; (b) Stackelberg equilibrium.}
\end{figure}

\subsection{The Stackelberg Equilibrium of Differential Game}
In the Stackelberg equilibrium issue, two players are considered, i.e., one leader and one follower. Different from the Nash equilibrium issue, the decisions of the players are not made simultaneously. The leader (\emph{player i}) makes the decision $u^i$ firstly. Based on the decision $u^i$, the follower (\emph{player j}) makes the decision $u^j$ to maximize his payoff function. For the follower, $\forall u^{i} \in U^i$, there exists the best decision set $\hbar (u^{i})$ to maximize the payoff $P^j (u^{i},u^{j})$, $ u^j \in \hbar (u^{i})$. In the Stackelberg equilibrium issue, we assume that the vehicle that has the smaller TCP is the leader, and the another vehicle is the follower.

If the following conditions hold, $(u^{i\ast},u^{j\ast})$ can be regarded as a Stackelberg equilibrium for the differential game.\\
(a) $ u^{j\ast} \in \hbar (u^{i\ast})$;\\
(b) $\forall u^{i} \in U^i$, and $ u^{j} \in U^j$
% Eq.
\begin{align}
P^i (u^{i},u^{j})\leq P^i (u^{i\ast},u^{j\ast})
\end{align}

According to the above definition of Stackelberg equilibrium, it can be found that the leader has to work out the best strategy of the follower with respect to his any decision $u^i$ firstly, and then makes the optimal decision $u^{i\ast}$ to maximize the payoff $P^i$.

Assuming that $(u^{i\ast},u^{j\ast})$ is a Stackelberg equilibrium, and $x^{\ast}(t)$ is the optimal state of the differential game system caused by  $(u^{i\ast},u^{j\ast})$, according to the PMP, there exists a adjoint vector $\zeta^j$ for the follower such that
% Eq.
\begin{align}
\left\{
\begin{array}{lr}
\dot{x}^{\ast}(t)=f(t, x^{\ast}(t),u^{i\ast}(t),u^{j\ast}(t))\\
\dot{\zeta}^j(t)=-\zeta^j \frac{\mathrm{\partial}f}{\mathrm{\partial}x} (t, x^{\ast}(t),u^{i\ast}(t),u^{j\ast}(t))+\\   \quad\quad\quad \quad \quad\quad\quad\quad \frac{\mathrm{\partial}J^j}{\mathrm{\partial}x} (t, x^{\ast}(t),u^{i\ast}(t),u^{j\ast}(t))\\
x(0)=x_0, \quad \zeta^j(T)=\nabla\phi^j(x(T))
\end{array}
\right.
\end{align}

Additionally,
% Eq.
\begin{align}
\begin{array}{lr}
u^{j\ast}(t)\in \arg\underset{\omega\in U^j}{\max} \{\zeta^j(t) f(t, x^{\ast}(t),u^{i\ast}(t),\omega)\\
\quad \quad \quad \quad \quad \quad -J^{j}(t,x^{\ast}(t),u^{i\ast}(t),\omega)\}\\
\end{array}
\end{align}

To find the optimal decision of the leader, the following assumption is proposed. For given $(t, x, u^i, \zeta^j)$, there exists a unique solution $u^{j\sharp}$ for the follower, i.e.,
% Eq.
\begin{align}
u^{j\sharp}(t,x,u^i,\zeta^j)\triangleq \arg\underset{\omega\in U^j}{\max} \{\zeta^j f(t, x,u^{i},\omega)-J^{j}(t,x,u^{i},\omega)\}
\end{align}

Furthermore, the optimal decision issue for the leader can be transformed into the following optimal control issue.
% Eq.
\begin{align}
u^{i\ast} = \arg\max P^i(t, x,u^i,u^{j\sharp}(t,x,u^i,\zeta^j))
\end{align}
subject to
% Eq.
\begin{align}
\left\{
\begin{array}{lr}
\dot{x}(t)=f(t, x,u^i,u^{j\sharp}(t,x,u^i,\zeta^j))\\
\dot{\zeta}^j(t)=-\zeta^j \frac{\mathrm{\partial}f}{\mathrm{\partial}x} (t, x,u^{i},u^{j\sharp}(t,x,u^i,\zeta^j))+\\   \quad\quad\quad \quad \quad\quad\quad\quad \frac{\mathrm{\partial}J^j}{\mathrm{\partial}x} (t, x,u^{i},u^{j\sharp}(t,x,u^i,\zeta^j))\\
x(0)=x_0, \quad \zeta^j(T)=\nabla\phi^j(x(T))
\end{array}
\right.
\end{align}

For the open-loop Stackelberg equilibrium solution, the two-player differential game issue is transformed into a single stage optimal control problem by Karush-Kuhn-Tucker (KKT) condition. The similar iterative solving process is conducted.
Fig. 3 shows the convergence analysis results of the algorithms. We can find that after 5 steps, the algorithms can converge to a stable value. The convergence rate meets the real-time implementation requirements for the decision-making algorithm.

\subsection{Decision-Making Process of AVs at Unsignalized Intersections}
In the game issue, the computational complexity increases with the rise of players. To reduce the computational burden of the hardware, not all AVs at the unsignalized intersection are always regarded as players in the differential game. It is controlled by a event trigger, which is associated with the collision risk. The event-triggered mechanism (ETM) is defined as follows.
% Eq.
\begin{align}
t_{k+1}\triangleq \inf \{t>t_k|\Gamma_{cp}^{ij}>\Gamma_{sf}\}
\end{align}
where $\Gamma_{cp}^{ij}$ denotes the total potential value of AV\emph{i} and AV\emph{j} at the predicted collision position. It means that when $\Gamma_{cp}^{ij}>\Gamma_{sf}$, the trigger condition is satisfied. As a result, it yields a differential game between AV\emph{i} and AV\emph{j}.

\begin{figure}[!b]\centering
	\includegraphics[width=5.5cm]{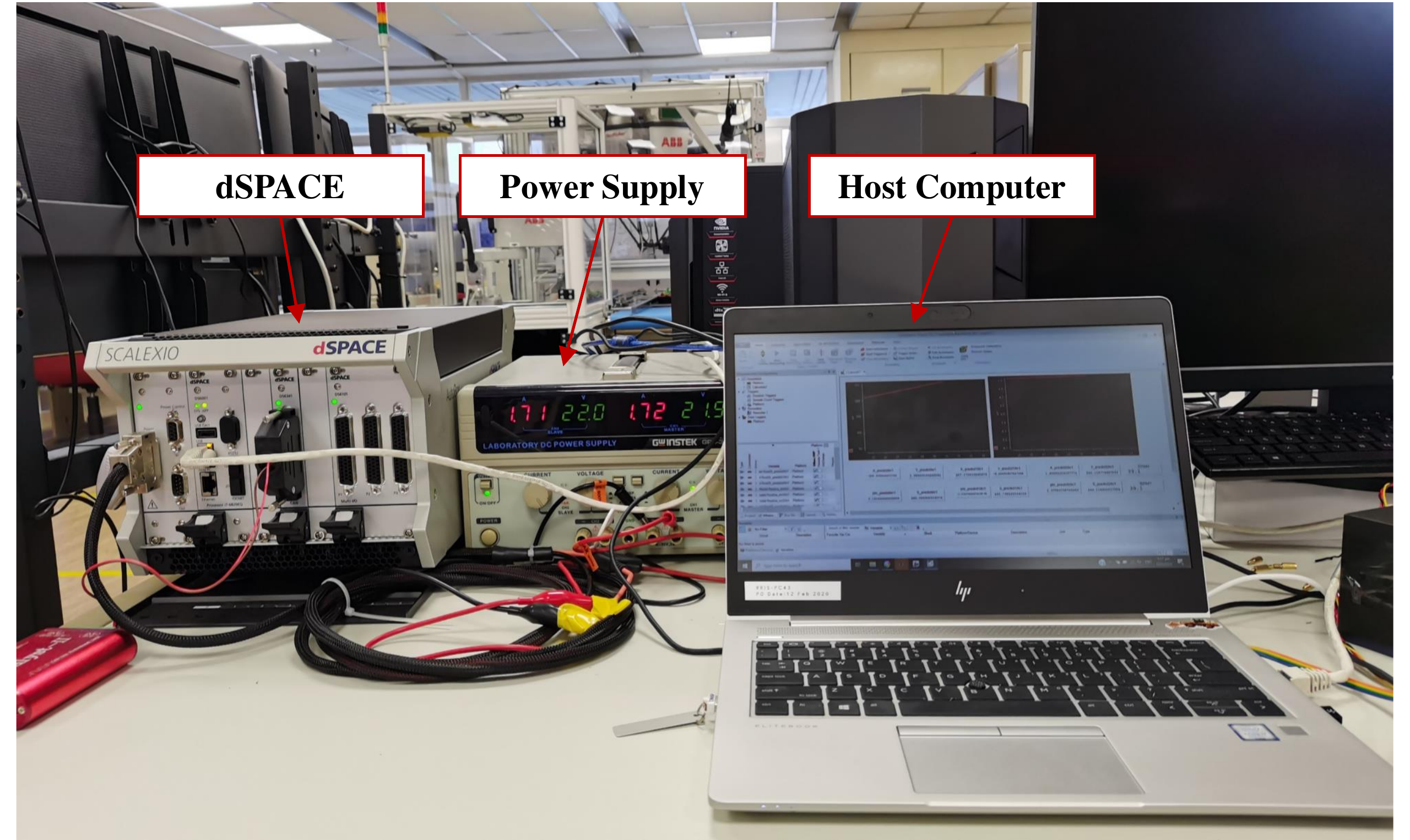}
	\caption{The HIL test platform for algorithm validation.}\label{FIG_3}
\end{figure}

Based on the proposed ETM, the decision-making process for AVs to pass the unsignalized intersection is designed in Algorithm 1.

\begin{algorithm}[h]
\caption{Decision making for AV\emph{i} to pass the unsignalized intersection.}
\begin{algorithmic}[1]
\FOR{$i=1:n$}
\STATE Input the motion state and position of AVs;\
\STATE Aggressiveness estimation of AVs;
\STATE Predict the CPs of AV\emph{i} with other vehicles;
\FOR{$j=1:m$}
\STATE Collision risk assessment of AV\emph{i} and AV\emph{j};
\IF {$\Gamma_{cp}^{ij}>\Gamma_{sf}$}
\STATE Conduct differential game between AV\emph{i} and AV\emph{j};
\STATE Calculate the Nash equilibrium and Stackelberg equilibrium;
\ELSE
\STATE No game between AV\emph{i} and AV\emph{j};
\STATE $k_{s-lat}^{i}=0$, $k_{s-lat}^{j}=0$;
\STATE Maximize $P^i$ and $P^j$ independently;
\ENDIF
\STATE Output the decision making results $u^{i*}$, $u^{j*}$;
\ENDFOR
\ENDFOR
\end{algorithmic}
\end{algorithm}

\begin{figure*}[!t]\centering
	\includegraphics[width=18cm]{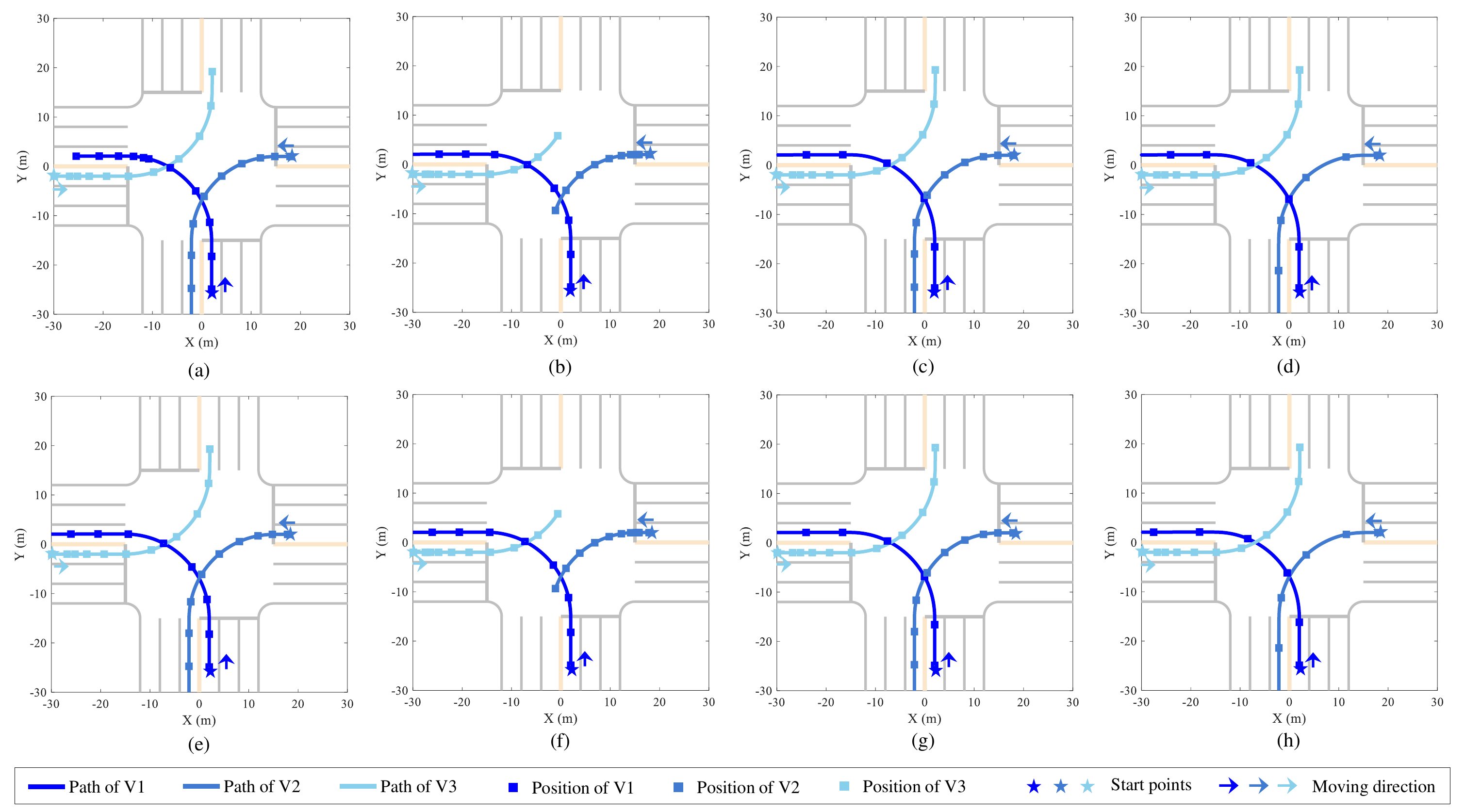}
	\caption{Decision-making results of AVs in Case 1: (a) NE in Scenario A; (b) NE in Scenario B; (c) NE in Scenario C; (d) NE in Scenario D; (e) SE in Scenario A; (f) SE in Scenario B; (g) SE in Scenario C; (h) SE in Scenario D.}\label{FIG_4}
\end{figure*}

\section{Hardware-in-the-loop Testing and Validation}
To evaluate the performance of the game theoretic-based decision making algorithm for AVs at unsignalized intersections, two test cases are carried out considering different preferences of AVs. All the testing scenarios and algorithm validation are implemented based on the hardware-in-the-loop (HIL) test platform.

\begin{table}[t]
	\renewcommand{\arraystretch}{1.2}
	\caption{Aggressiveness of AVs in Case 1}
\setlength{\tabcolsep}{2.5 mm}
	\centering
	\label{table_3}
	%\centering
	\resizebox{\columnwidth}{!}{
		\begin{tabular}{c c c c c}
			\hline\hline \\[-4mm]

\multicolumn{1}{c}{Aggressiveness  $\kappa$} & Scenario A & Scenario B & Scenario C & Scenario D \\
\hline
			\multicolumn{1}{c}{V1} & 0.5 & 0.5 & 0.8 & 0.8 \\
			\multicolumn{1}{c}{V2} & 0.5 & 0.2 & 0.5 & 0.9 \\
  \multicolumn{1}{c}{V3} & 0.5 & 0.2 & 0.5 & 0.5 \\

			\hline\hline
		\end{tabular}
	}
\end{table}

\subsection{The HIL Testing Platform}
A hardware system, i.e. the dSPACE SCALEXIO Autobox, is used as the algorithm validation platform to evaluate the effectiveness and real-time performance of the proposed decision-making algorithm. The HIL test platform is shown in Fig. 4, which mainly includes three parts, i.e., SCALEXIO Autobox, power supply and host computer. Software platform is made up of three parts, i.e., Matlab/Simulink, ConfigurationDesk, and ControlDesk. The work process of the HIL test platform is described as follows. Firstly, the decision-making algorithm in Matlab/Simulink is built to C code via ConfigurationDesk software. Then, the C code is downloaded into the SCALEXIO AutoBox hardware. Through the ControlDesk software, we can monitor and record the test data on the host computer.

\subsection{Validation and Results}
To evaluate the performance of the proposed decision-making algorithm, two HIL test cases are designed and conducted. In Case 1, different driving aggressiveness coefficients are set for AVs to evaluate the effect of personalized driving preferences on decision making. In Case 2, more vehicles and more complex driving conflicts are considered. Moreover, the proposed two equilibrium solutions to the differential game approach, i.e., Nash equilibrium (NE) and Stackelberg equilibrium (SE), are verified in two cases.

(1) Case 1

\begin{table}[!t]
	\renewcommand{\arraystretch}{1.2}
	\caption{Minimum Time to Collision in Case 1}
\setlength{\tabcolsep}{1 mm}
	\centering
	\label{table_5}
	%\centering
	\resizebox{\columnwidth}{!}{
		\begin{tabular}{c c c | c c | c c | c c c c c c }
			\hline\hline \\[-4mm]
			\multirow{2}{*}{TTC Min / s} & \multicolumn{2}{c|}{Scenario A} & \multicolumn{2}{c|}{Scenario B} & \multicolumn{2}{c|}{Scenario C} & \multicolumn{2}{c}{Scenario D} \\
\cline{2-9} & \makecell [c] {V1-V2} & \makecell [c] {V1-V3} & \makecell [c] {V1-V2} & \makecell [c] {V1-V3} & \makecell [c] {V1-V2} & \makecell [c] {V1-V3} & \makecell [c] {V1-V2} & \makecell [c] {V1-V3} \\
\hline
			\multicolumn{1}{c}{NE} & 3.71 & 3.57 & 43.62 & 10.15 & 5.37 & 6.81 & 0.70 & 6.87 \\
            \multicolumn{1}{c}{SE} & 3.82 & 3.87 & 46.58 & 10.60 & 5.43 & 6.85 & 0.80 & 7.15 \\

			\hline\hline
		\end{tabular}
	}
\end{table}

\begin{figure}[!t]\centering
	\includegraphics[width=8cm]{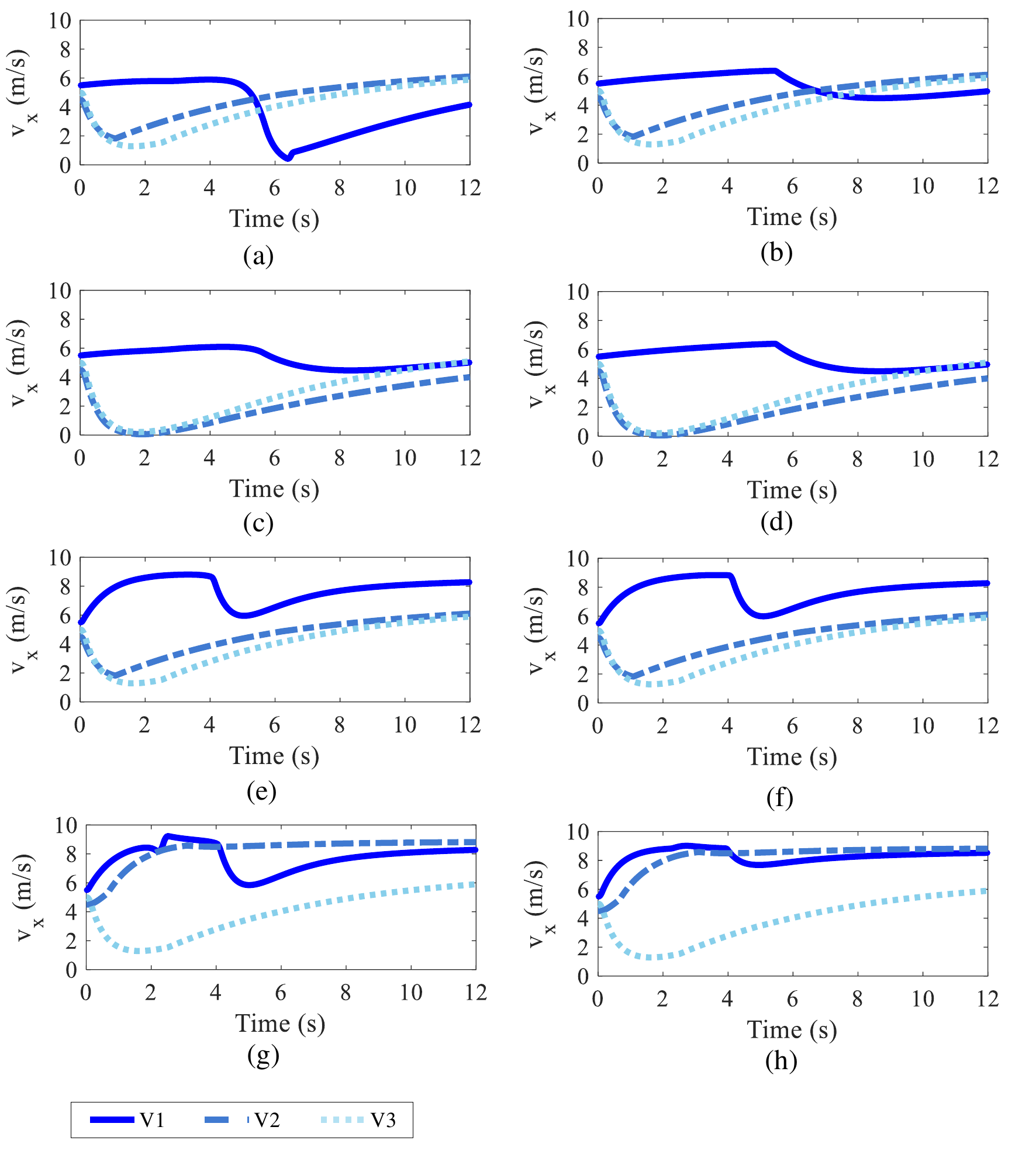}
	\caption{Velocities of AVs in Case 1: (a) NE in Scenario A; (b) SE in Scenario A; (c) NE in Scenario B; (d) SE in Scenario B; (e) NE in Scenario C; (f) SE in Scenario C; (g) NE in Scenario D; (h) SE in Scenario D.}\label{FIG_5}
\end{figure}

\begin{figure}[!t]\centering
	\includegraphics[width=8cm]{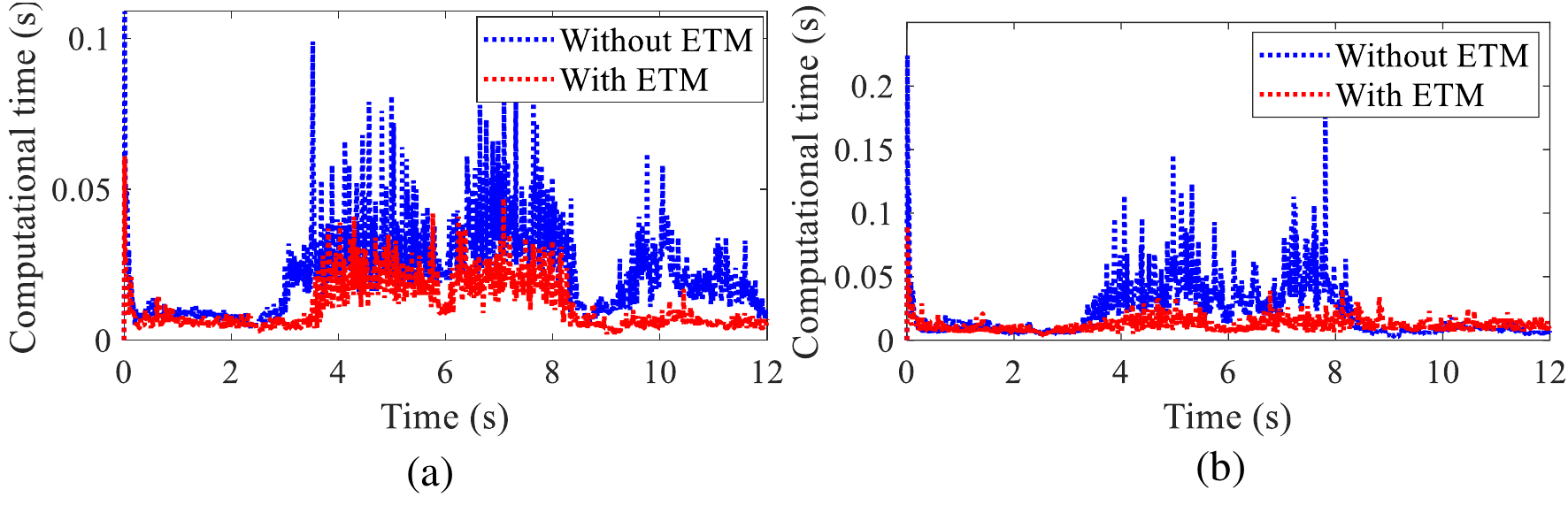}
	\caption{Computational efficiency: (a) NE; (b) SE.}\label{FIG_6}
\end{figure}

In this case, three AVs are considered at the unsignalized intersection. Four typical scenarios are designed according to different aggressiveness combinations of the three AVs, which are shown in Table I. In this case, all the three AVs want to turn left, which yields two cross conflicts. The start point positions of V1, V2 and V3 are set as (2, -25), (18, 2) and (-30, -2), respectively. In addition, the initial velocities of V1, V2 and V3 are 5.5 m/s, 4.5 m/s and 5 m/s, respectively.

The moving trajectories and velocities of AVs in Case 1 are illustrated in Figs. 4 and 5, respectively. In Scenario A, it can be found from Fig. 5 (a), (e) and Fig. 6 (a), (b) that both V2 and V3 decelerated and gave ways for V1. In both NE and SE, the velocity curves of V2 and V3 fall firstly and then rise. In the interaction and game process with V1, V2 and V3 had to decelerate to guarantee driving safety. After passing the conflict points, V2 and V3 accelerated to improve travel efficiency. Compared with Scenario A, the aggressiveness values of V2 and V3 are reduced in Scenario B. As a result, the position points of V2 and V3 in Fig. 5 (b) and (f) are denser before passing the conflict points, and the velocity curves of V2 and V3 in Fig. 6 (c) and (d) fall sharply at first, which indicates the low aggressiveness brings better driving safety at the expense of traffic efficiency. V2 and V2 in Scenario B are just like the cautious human drivers. In Scenario C, the aggressiveness value of V1 is increased compared with that in Scenario A. From Fig. 6 (e) and (f), we can find that the velocities of V2 and V3 have no obvious change. However, V1's velocity is increased remarkably, which means the travel efficiency is improved with the rise of driving aggressiveness. In Scenario D, the aggressiveness value of V2 is increased compared with Scenario C. Both V1 and V2 are very aggressive, yielding the high passing velocity in Fig. 6 (g) and (h). Due to the high aggressiveness, the position points of V1 and V2 in Fig. 5 (d) and (h) are very close before passing the conflict point, which is harmful to driving safety. Besides, for the detailed analysis of driving safety, the TTC results between conflict vehicles are listed in Table II. It can be concluded that TTC decreases with the rise of aggressiveness coefficients of vehicles, which means worse driving safety, especially when the vehicles are all highly aggressive, e.g., V1 and V2 in Scenario D.

Additionally, it can be found from Figs. 5 and 6 that NE and SE have similar decision-making results on driving behaviors. Compared with the test results in NE, there is no obvious velocity change for the follower players in SE. For instance, the velocity curves of both V2 and V3 in Fig. 6 (a) and (b) are very similar. However, the leader player always adjusts its strategy to improve both driving safety and passing efficiency thanks to the power of predicting follower's strategy given its own. Especially when the leader player vehicle has the similar aggressiveness coefficient to its competitor, the velocity adjustment is very obvious, e.g., V1 in Scenario A and Scenario D. In general, both NE and SE can provide the personalized and human-like decision making for AVs at unsignalized intersections.

To verify the effect of the ETM method on the computational efficiency of the decision-making algorithm, the comparative test is conducted. Taking Scenario A as an instance, the test results of computational efficiency are illustrated in Fig. 7. For the NE algorithm, the mean values of computational time with ETM and without ETM for each sampling step are 0.0116s and 0.0226s, respectively. The computational time is reduced by $48\%$ with ETM. For the SE algorithm, the mean values of computational time with ETM and without ETM for each sampling step are 0.0129s and 0.0241s, respectively. The computational time is reduced by $46\%$ with ETM. It can be concluded that the proposed ETM can improve the computational efficiency for the decision-making algorithm, reducing the computational burden for the hardware platform.

(2) Case 2

In Case 2, the driving conflicts of five AVs at the unsignalized intersection are discussed. Two scenarios considering different aggressiveness combinations are designed, which are shown in Table III.
In this case, V1 on the middle lane wants to pass the intersection from west to east. V2 on the middle lane wants to pass the intersection from north to south. V3 and V4 on the left lane want to turn left from north to east. V5 on the right lane wants to turn right from south to east. Three kinds of driving conflicts exist in this case, i.e., cross conflict between V1 and V2, following conflict between V3 and V4, confluence conflicts between V1 and V3, V1 and V5. The start point positions of V1, V2, V3, V4 and V5 are set as (-28, -6), (-6, 25), (-2, 16), (4, 2) and (10, -50), respectively. The initial velocities of V1, V2, V3, V4 and V5 are 6 m/s, 5 m/s, 5.5 m/s, 5 m/s and 6 m/s, respectively.

The testing results of Case 2 are illustrated in Figs. 8 and 9. Fig. 8 (a) and (b) shows the decision-making results of AVs in Scenario A and Scenario B with the NE approach. Due to the different settings of AVs' driving aggressiveness in Scenario A and Scenario B, it yields different decision making results. It can be found from Fig. 8 (a) and (b) that V3's moving trajectories are quite different in two scenarios. In Scenario A, due to the large aggressiveness, V3 chose to accelerate and merge into the middle lane. The velocity curve of V3 in Fig. 9 (a) keeps rising until the maximum vale. However, the aggressiveness of V3 is reduced in Scenario A. As a result, the position points of V3 are very dense in Fig. 8 (b), and the velocity curve of V3 in Fig. 9 (b) has a sharp drop in the process of passing the conflict zone. Namely, V3 made the decision that decelerating and merging into the left lane. Besides, due to the different settings of aggressiveness, V1 also has some obvious decision-making differences in the two scenarios. The velocity curve of V1 has different changes in Fig. 9 (a) and (b). The decision-making differences of other AVs are relatively small. As to the conflict resolution, the driving aggressiveness of V1 is larger than V2 in the two scenarios. As a result, the cross conflict between V1 and V2 is solved with the accelerating of V1 and the decelerating of V2. In Scenario A, due to the low speed of V4, V3 gave up car following and turned left to the middle lane rather than the left lane. As a result, a driving conflict between V3 and V1 is unavoidable. Since V3 is more aggressive than V1, V1 chose to slow down and gave way to V3. In Scenario B, due to the small driving aggressiveness of V3, i.e., conservative driving style, V3 made the different decision, slowing down and following V4. V1 passed the intersection with high speed. According to the results of Figs. 8 and 9, we can find that the driving aggressiveness larger, the vehicle velocity larger, which verifies the aggressive driving style prefers high passing efficiency. For the test results of the SE approach, the similar conclusion can be conducted.

Although both NE and SE can help AVs address driving conflicts and make safe decisions at unsignalized intersections, there exist some decision-making differences between them. In SE, before the the leader vehicle makes decisions, the decision-making result of follower vehicle is predicted and considered in advance by the leader vehicle. As a result, the leader vehicle can adjust its strategy to improve both driving safety and passing efficiency. The detailed analysis in Tables IV supports the conclusion.

\begin{table}[t]
	\renewcommand{\arraystretch}{1.2}
	\caption{Aggressiveness of AVs in Case 2}
\setlength{\tabcolsep}{4.5 mm}
	\centering
	\label{table_3}
	%\centering
	\resizebox{\columnwidth}{!}{
		\begin{tabular}{c c c c c c}
			\hline\hline \\[-4mm]

			\multicolumn{1}{c}{Aggressiveness  $\kappa$} & V1 & V2 & V3 & V4 & V5 \\
\hline
			\multicolumn{1}{c}{Scenario A} & 0.50 & 0.40 & 0.85 & 0.45 & 0.50 \\
            \multicolumn{1}{c}{Scenario B} & 0.90 & 0.40 & 0.20 & 0.45 & 0.50 \\

			\hline\hline
		\end{tabular}
	}
\end{table}

\begin{table}[!t]
	\renewcommand{\arraystretch}{1.2}
	\caption{Minimum Time to Collision in Case 2}
\setlength{\tabcolsep}{2.2 mm}
	\centering
	\label{table_5}
	%\centering
	\resizebox{\columnwidth}{!}{
		\begin{tabular}{c c c c | c c c }
			\hline\hline \\[-4mm]
			\multirow{2}{*}{TTC Min / s} & \multicolumn{3}{c|}{Scenario A} & \multicolumn{3}{c}{Scenario B} \\
\cline{2-7} & \makecell [c] {V1-V2} & \makecell [c] {V1-V3} & \makecell [c] {V3-V4} & \makecell [c] {V1-V2} & \makecell [c] {V1-V3} & \makecell [c] {V3-V4} \\
\hline
			\multicolumn{1}{c}{NE} & 2.53 & 2.39 & 30.39 & 5.74 & / & 18.42 \\
            \multicolumn{1}{c}{SE} & 3.93 & 2.50 & 30.39 & 6.28 & / & 19.04 \\

			\hline\hline
		\end{tabular}
	}
\end{table}

\begin{figure}[!t]\centering
	\includegraphics[width=8cm]{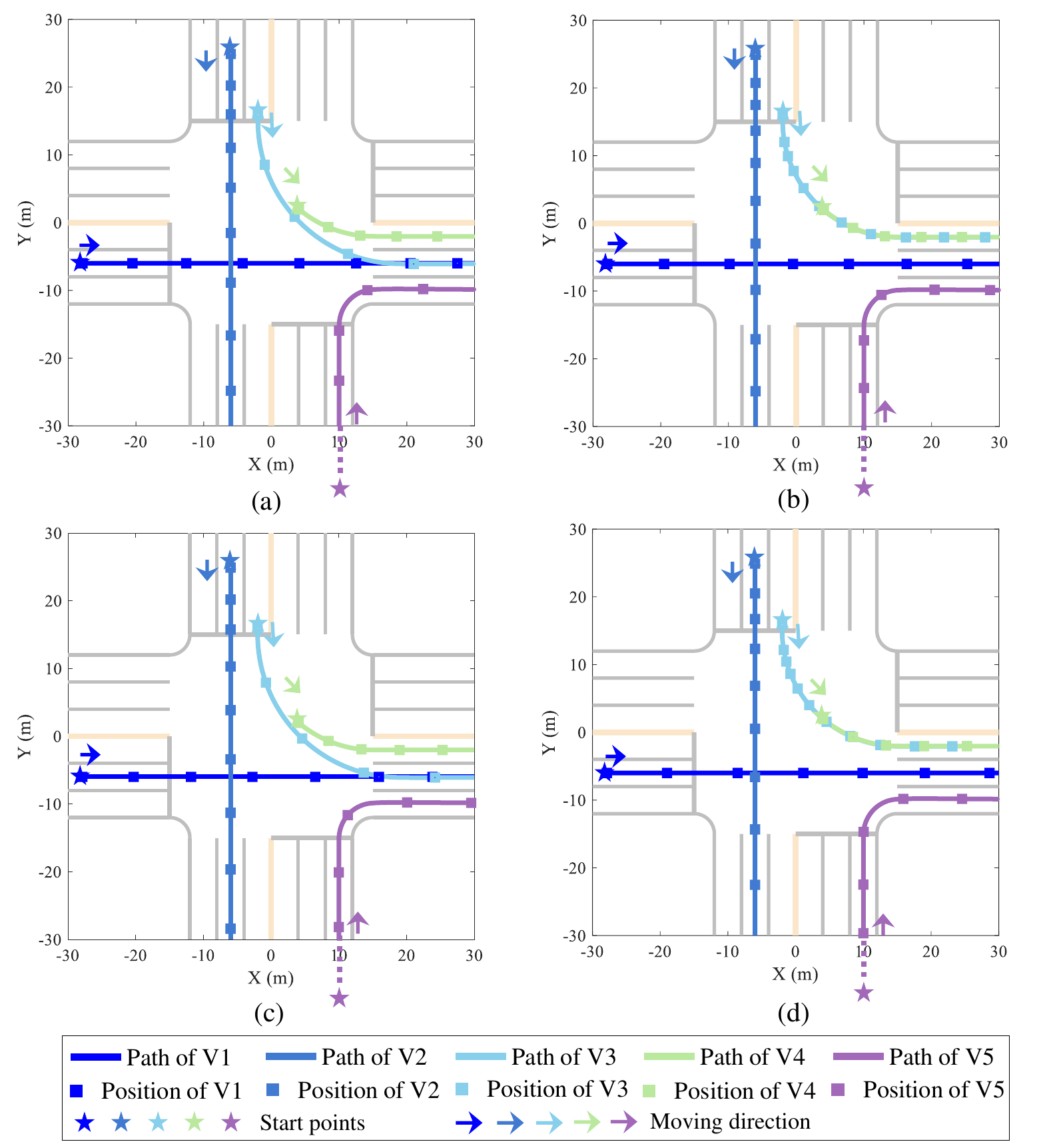}
	\caption{Decision-making results of AVs in Case 2: (a) NE in Scenario A; (b) NE in Scenario B; (c) SE in Scenario A; (d) SE in Scenario B.}\label{FIG_7}
\end{figure}

\begin{figure}[!t]\centering
	\includegraphics[width=8cm]{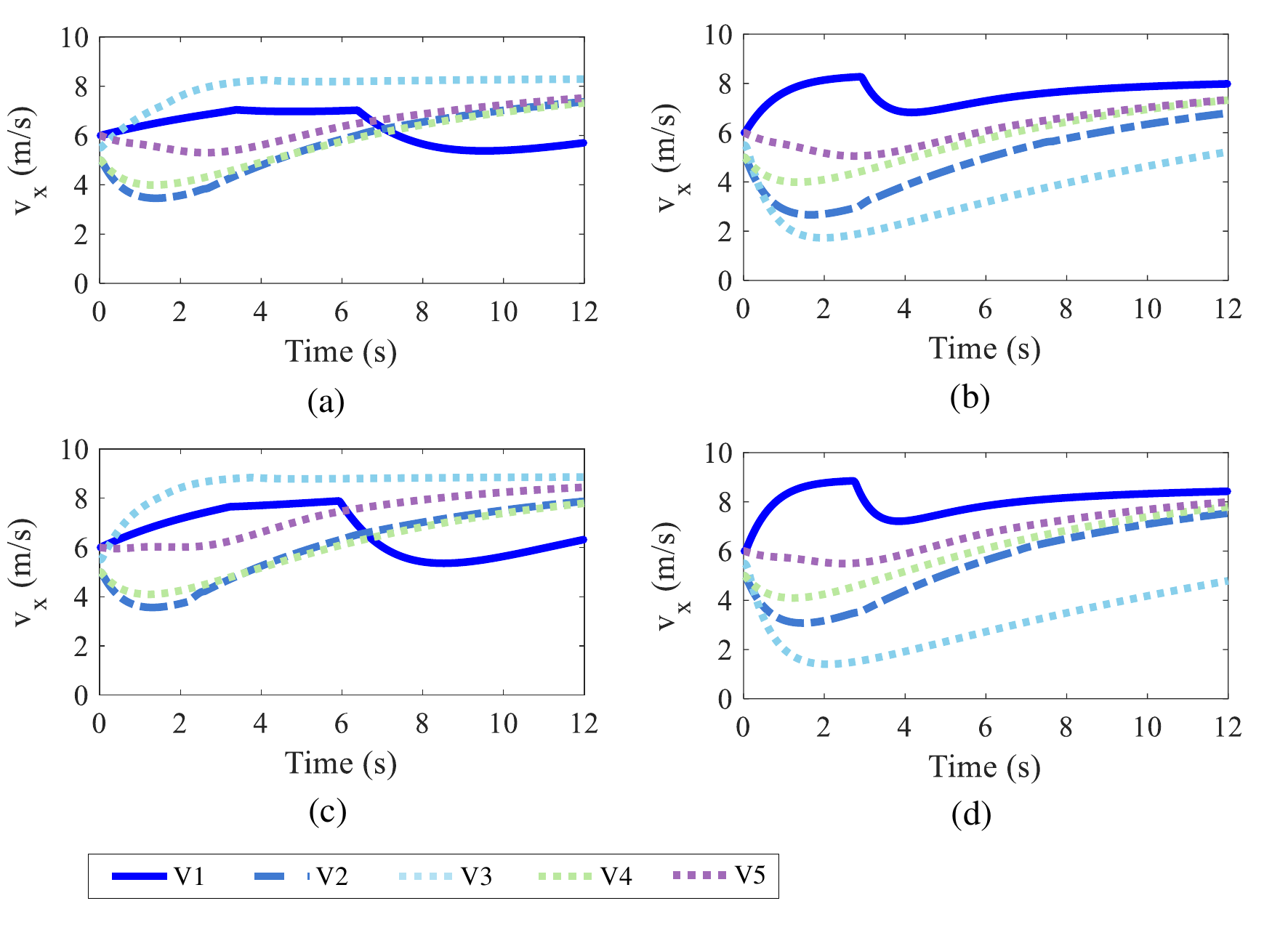}
	\caption{Velocities of AVs in Case 2: (a) NE in Scenario A; (b) NE in Scenario B; (c) SE in Scenario A; (d) SE in Scenario B.}\label{FIG_8}
\end{figure}

\subsection{Discussion}
From the test results of two cases, it can be concluded that the proposed decision-making algorithm can address all kinds of driving conflicts for AVs at unsignalized intersections. Additionally, personalized decision-making results of AVs are reflected by the setting of aggressiveness parameters. Large aggressiveness parameter is beneficial to the improvement of passing efficiency, but will worsen the driving safety especially in the condition of existing two aggressive AVs. Small aggressiveness parameter is in favor of the safety advancement for AVs, but will decrease the passing efficiency. For future AV design, the personalized decision making of AVs can be realized by defining the aggressiveness level.

\section{Conclusion}
Based on game theory, a novel decision making framework for AVs is proposed to address driving conflicts at unsignalized intersections. The personalized driving preferences of AVs are taken into account in the decision-making framework. Moreover, the collision risk assessment algorithm is designed with a Gaussian potential field model in favor of the performance advancement of the decision-making framework. In the payoff function construction of decision making, driving safety, passing efficiency and driving aggressiveness are considered. Then, a differential game approach is applied to the driving conflict resolution of AVs. Both Nash equilibrium and Stackelberg equilibrium of the differential game are discussed and solved. To reduce the computational complexity of the decision-making algorithm, an event-triggered mechanism is designed. Finally, the proposed method is tested under two representative driving cases on the HIL platform. Based on the testing results, it can be concluded that the proposed decision-making algorithm can make safe and efficient decisions for AVs, meanwhile it can guarantee the personalized driving preferences, demonstrating its feasibility, effectiveness and real-time implementation performance.

Our future work will focus on the decision-making issue of AVs in the human-machine hybrid driving environment. Scene visualization in the HIL test will be considered as well. Besides, cyber-security is a critical issue in autonomous driving [42]. Cyber-attack will be considered in the decision-making process of AVs, for further improving driving safety.

\end{document}